\documentclass[aps,prd,groupedaddress,showpacs,showkeys]{revtex4}
\usepackage{epsfig}
\pdfoutput=1
\newcommand{\be}[1]{\begin{equation} \label{(#1)}}
\newcommand{\eq}{\begin{eqnarray}}
\newcommand{\ee}{\end{equation}}
\newcommand{\en}{\end{eqnarray}}
\newcommand{\ba}[1]{\begin{eqnarray} \label{(#1)}}
\newcommand{\ea}{\end{eqnarray}}
\newcommand{\nn}{\nonumber}
\newcommand{\rf}[1]{(\ref{(#1)})}
\begin{document}

%\title{Model Independent Limits on Sterile neutrino mixing with $\nu_{\tau}$}
\title{On sterile neutrino mixing with $\nu_{\tau}$}
\author{Juan Carlos Helo,  Sergey Kovalenko, Ivan Schmidt  \vspace*{0.3\baselineskip}}
\affiliation{Departamento de F\'\i sica, Universidad T\'ecnica Federico Santa Mar\'\i a, \\
and\\
 Centro-Cient\'\i fico-Tecnol\'{o}gico de Valpara\'\i so, \\
Casilla 110-V, Valpara\'\i so,  Chile \vspace*{0.3\baselineskip}\\}

\date{\today}

\begin{abstract}
Matrix element $U_{\tau N}$ of sterile neutrino $N$ mixing with $\nu_{\tau}$ is the least constrained 
in the literature among the three $U_{\alpha N}$ ($\alpha=e,\mu,\tau$) mixing parameters 
characterizing the sterile neutrino phenomenology. 
We study the contribution of massive dominantly sterile neutrinos to
purely leptonic $\tau$-decays and semileptonic decays of $\tau$ and K, D mesons. We consider some decays allowed in the Standard Model (SM) as well as Lepton Flavor and Lepton Number Violating (LFV, LNV) decays forbidden in the SM.
From the existing experimental data on the branching ratios of these processes we derived new limits on 
$U_{\tau N}$ more stringent than the ones existing in the literature. 
These limits are extracted in a model independent way without any 
ad hoc assumptions on the relative size of the three different sterile neutrino mixing parameters. 
\end{abstract}

\pacs{13.35.Hb, 13.15.+g, 13.20.-v, 13.35.Dx}

%{12.15.-y,12.60.-i,13.15.+g,13.20.-v,14.60.Pq}

\keywords{sterile neutrino, tau-lepton, lepton number and lepton flavor violation}

\maketitle

\newpage

\section{Introduction}
\label{sec-1}

Lepton flavors are conserved in the Standard Model (SM) due to the presence of an  accidental lepton flavor
%global $U_{1e}\times U_{1\mu}\times U_{1\tau}$ 
symmetry, which, however, is broken by non-zero neutrino masses. 
Neutrino oscillation experiments have proven that neutrinos are massive, although very light, particles mixing with each other. 
Moreover, neutrino oscillations is the first and so far  the only observed phenomenon of lepton flavor violation (LFV). 
In the sector of charged leptons LVF is strongly suppressed by the smallness of neutrino square mass differences 
$(m^{2}_{\nu_{i}}-m^{2}_{\nu_{j}})/q^{2}_{0}$ compared to the characteristic momentum scale, $q_{0}$, of an LFV process
which is typically of the order of the charged lepton mass $q_{0}\sim m_{l}$.
If neutrinos are Majorana particles there can also occur lepton number violating (LNV) processes. They are also suppressed by the smallness of the absolute value of $m_{\nu}$. 
However, the situation may dramatically change if there exist either heavy neutrinos $N_{i}$, known as sterile, mixed with 
the active flavors $\nu_{e,\mu,\tau}$ or if there are some new LFV and LNV interactions beyond the SM. 

Here we study the former possibility and consider an extension of the SM with  right-handed neutrinos.
In the case of  $n$ species of the SM singlet right-handed neutrinos
$\nu^{\,\prime}_{Rj}=(\nu^{\,\prime}_{R1},...\nu^{\,\prime}_{Rn})$,
besides the three left-handed weak doublet neutrinos
$\nu^{\,\prime}_{Li} =
(\nu^{\,\prime}_{Le},\nu^{\,\prime}_{L\mu},\nu^{\,\prime}_{L\tau})$
the neutrino mass term can be written as 
\begin{eqnarray}\label{Mass-Term}
-\frac{1}{2} \overline{\nu^{\,\prime}} {\cal M}^{(\nu)} \nu^{\,\prime
c} + \mbox{h.c.} & = & - \frac{1}{2} (\bar\nu^{\,\prime}_{_L},
\overline{\nu_{_R}^{\,\prime c}}) \left(\begin{array}{cc}
{\cal M}_L & {\cal M}_D \\
{\cal M}^T_D  & {\cal M}_R \end{array}\right) \left(\begin{array}{c}
\nu_{_L}^{\,\prime c} \\
\nu^{\,\prime}_{_R}\end{array}\right) + \mbox{h.c.} \\
&=&-\frac{1}{2} (\sum_{i=1}^{3} m_{\nu_i} \overline{\nu^c_i}\nu_i
+\sum_{j=1}^{n} m_{\nu_j} \overline{\nu^c_j}\nu_j  )+ \mbox{h.c.}
\end{eqnarray}
Here ${\cal M}_L, {\cal M}_R$ are $3\times 3$ and $n\times n$
symmetric Majorana mass matrices, and ${M}_D$ is a $3\times n$ Dirac
type matrix. Rotating the neutrino mass matrix to the diagonal form by a unitary
transformation
 \begin{eqnarray}\label{mixing-matrix}
 && U^T {\cal M}^{(\nu)}U =
\textrm{Diag}\{m_{\nu 1}, \cdots ,m_{\nu_{3+n}}\}
\label{mass-eigen}
\end{eqnarray}
one ends up with $3+n$
Majorana neutrinos with masses $m_{v_1}, \cdots, m_{v_{3+n}}$.  The matrix $U_{\alpha k}$ is a neutrino mixing matrix.
%%%%%%%%%%%%%%%%%%%%%
In special cases among neutrino mass eigenstates there may appear  pairs with masses degenerate in
absolute values. Each of these pairs can be collected into a Dirac neutrino
field. This situation corresponds to conservation of certain lepton numbers
assigned to these Dirac fields. Generically in this setup neutrino mass eigenstates can be of any mass. 
For consistency with neutrino phenomenology (for recent review, cf.  \cite{rev-nu-phen}) among them there must be the three very light neutrinos with different masses and  dominated by the active flavors $\nu_{\alpha}$ ($\alpha = e, \mu, \tau$).  The remaining states may also have certain admixture of the active flavors and, therefore, participate in charged and neutral current interactions of the SM contributing to LNV and LFV processes. 
Explanation of the presence in the neutrino spectrum of  the three very light neutrinos requires additional physically motivated assumptions on the structure of the mass matrix in (\ref{Mass-Term}). The celebrated ``see-saw'' mechanism \cite{see-saw}, presently called type-I see-saw, is implemented in this framework assuming that  $\mathcal{M}_R\gg \mathcal{M}_D$. Then,  there naturally appear light neutrinos with masses of the order of $\sim\mathcal{M}_D^2/\mathcal{M}_R$ dominated by $\nu_{\alpha}$. Also, there must be present heavy Majorana neutrinos with masses at the scale of  $\sim \mathcal{M}_R$. Their mixing with active neutrino flavors is suppressed by a factor $\sim \mathcal{M}_D/\mathcal{M}_R$ which should be very small. In particular scenarios this generic limitation of the see-saw mechanism can be relaxed \cite{relax}. Then the heavy neutrinos could be, in principle, observable at LHC, if their masses are within the kinematical reach the corresponding experiments. Very heavy or moderately heavy Majorana entry 
$\mathcal{M}_R$ of the the neutrino mass matrix naturally appears in various extensions of the SM. The well known examples are given by the $SO(10)$-based supersymmetric \cite{SUSY-GUT} and  ordinary \cite{GUT} grand unification models.
The supersymmetric versions of see-saw  are also widely discussed in the literature 
(see, for instance,  \cite{SUSY-see-saw} and references therein). 
%  (for recent review see, for instance, Ref. \cite{rev-nu-mass}).  

%
In the present paper we study the above mentioned generic case of the neutrino mass matrix in (\ref{Mass-Term}) without implying a specific scenario of neutrino mass generation.  
We assume there is at least one moderately heavy neutrino $N$ in the MeV-GeV domain or even lighter.
%, which we are going to consider in the present paper.
The presence or absence of these neutrino states, conventionally called sterile neutrinos, is a question for experimental searches.
If exist, they may contribute to some LNV and LFV processes as intermediate nearly on-mass-shell states. This would lead to resonant enhancement of their contributions to these processes.  As a result, it may become possible to either observe the LNV, LFV processes or  set stringent limits on sterile neutrino mass $m_{N}$  and mixing $U_{\alpha N}$ with active neutrino flavors $\nu_{\alpha}$ ($\alpha = e, \mu, \tau$) from non-observation of the corresponding processes.

On the other hand the sterile neutrinos in this mass range are motivated by various phenomenological models \cite{Mohapatra}, in particular, by the recently proposed electroweak scale see-saw models \cite{EWSS1},  \cite{EWSS2}.  
They may also play an important astrophysical and cosmological role.  The sterile neutrinos in this mass range may have an impact on Big Bang nucleosynthesis, large scale structure formation \cite{nuclsyn}, supernovae explosions \cite{supernovae}. Moreover, the keV-GeV sterile neutrinos are good dark matter candidates \cite{DM-Bar-1,DM-Bar-2,DM-Bar-3}  and offer a
plausible explanation of  baryogenesis \cite{Barg}.
Dark Matter sterile neutrinos, having small admixture of active flavors, may suffer radiative decays and contribute to the diffuse extragalactic radiation and x-rays from galactic clusters \cite{galclast}. 
This is, of course, an incomplete list of cosmological and astrophysical  implications of sterile neutrinos.  More details on this subject can be found  in  Refs. \cite{Dolgov1}, \cite{Smirnov1}.

The phenomenology of sterile neutrinos in the processes, which can be searched for in laboratory experiments have been studied in the literature in different contexts and from complementary points of view (for earlier studies see \cite{Shrock}).
Their resonant contributions to $\tau$ and meson decays have been studied in Refs.  
\cite{K-decPaper, tau-decPaper,Ivanov:2004ch, M-decPaper,Cvetic:2010rw,Atre:2009rg}. 
Another potential process to look for  sterile Majorana neutrinos is like-sign dilepton production in hadron collisions  
\cite{Almeida:2000pz,Panella:2001wq,Han:2006ip,Kovalenko:2009td}.  Possible implications of sterile neutrinos have been also studied in LFV muonium decay and high-energy muon-electron scattering \cite{Cvetic:2006yg}.
An interesting explanation of anomalous excess of events observed in the LSND \cite{LSND} and MiniBooNE \cite{MiniBooNE} neutrino  experiments has been recently proposed  \cite{Gninenko} in terms of  sterile neutrinos with masses from 40 MeV to 80 MeV.  An explanation comes out of their possible production in neutral current interactions of $\nu_{\mu}$ and subsequent radiative decay  to light neutrinos.

Here we study  a scenario with only one sterile neutrino state $N$.
% with mass $m_{N}$ in the MeV-GeV region.
% and examine their phenomenological impact on the LFV decays of $\tau$ and some mesons.  
Phenomenology of a single sterile neutrino $N$
is specified by its mass $m_{N}$ and three mixing matrix elements  $U_{eN}$, $U_{\mu N}$, $U_{\tau N}$.
% characterizing sterile neutrino mixing with active SM flavors. 
%Previously in Ref. \cite{M-decPaper}, analyzing semileptonic LFV and LNV decays  of   $\tau$ and some light and heavy %mesons, we extracted from the corresponding experimental data upper limits on $U_{eN}, U_{\mu N}$. 
In the present paper we focus on the derivation of limits on the matrix element $U_{\tau N}$, which is currently least constrained in the literature.
Towards this end we use the results of experimental measurements of branching ratios of purely leptonic 
$\tau$ decays and semileptonic decays of $\tau$ and $K, D$ mesons \cite{PDG}. One of the key points of our derivation is its model independent character,
in the sense that we do not apply any additional assumptions on the relative size of the three mixing parameters 
$U_{\alpha N}$. Such ad hoc assumptions are typical in the literature and stem from the fact that all these 
three parameters enter in the decay rate formulas of any decay, potentially receiving contribution from $N$ as an intermediate state. 
Therefore, in order to extract individual limits on each mixing parameter one may need additional information on them.
We will show that in purely leptonic $\tau$ decays it is unnecessary and in the other cases this sort of information can be procured by a joint analysis of certain sets of leptonic and semileptonic decays of $\tau$ and $K, D$.  
%which we combine 
%$\tau^{-}\rightarrow e^{-}e^{-}e^{+}\nu\nu$ and 
%$\tau^{-}\rightarrow \mu^{-}e^{-}e^{+}\nu\nu$ decays
%\cite{PDG}.  
%In certain domains of sterile neutrino mass $m_{N}$, in order to extract model independent limits on $U_{\tau N}$, we combine %these
%decays with semileptonic decays of $\tau$ and some mesons.
%
%$\tau^{-} \rightarrow \pi^{-}\pi^{\pm}l^{\mp}$,  $M^{+}\rightarrow \pi^{-}\mu^{+}e^{+}$, where $M=K, D, B$ mesons.

The paper is organized as follows. 
In the next Section \ref{sec-2} we present decay rate formulas for $\tau$ and pseudoscalar meson LFV and LNV decays in the resonant domains of sterile neutrino mass $m_{N}$.
In Section \ref{sec-3} we derive upper limits on $|U_{\tau N}|$ from the existing experimental data on purely leptonic 5-body $\tau$ 
decays, semileptonic $\tau$ and $K, D$ decays, considering sterile neutrino contribution as an intermediate state and in some cases as one of the final state particles.   Section \ref{sec-4} contains summary and discussion of our main results.

\section{ Decay Rates}
\label{sec-2}

Neutrino interactions are represented by the SM Charged (CC) and Neutral Current (NC) Lagrangian terms.  In the mass eigenstate basis they read
\begin{eqnarray}\label{CC-NC}
{\cal L} = \frac{g_2}{\sqrt{2}}\sum_{i}\  U_{l i}\ \bar l \gamma^{\mu} P_L \nu_i\  W^-_{\mu}
+  \frac{g_2}{2 \cos\theta_W}\  \sum_{\alpha, i, j}U_{\alpha j} U_{\alpha i}^*\  \bar\nu_i \gamma^{\mu} P_L \nu_j\  
Z_{\mu},
\end{eqnarray}
where $l = e,\mu,\tau$ and $i=1,..., n+3 $. We consider the case with a single sterile neutrino $N$ and, therefore, we choose $n=1$ and identify $N=\nu_{4}$.

In what follows we study sterile neutrino contribution to the following decays
\begin{eqnarray}\label{DECAYS}
 \tau^- &\rightarrow& l^{-} e^{-} e^+ \nu\  \nu,\ \  \ \ \tau^-\rightarrow l^{\mp} \pi^{\pm} \pi^-, \ \  \ \ M^+\rightarrow l_1^{+} l_2^{\pm} \pi^{\mp}\\
 \label{TAU-N-th} 
 \tau^{-} &\rightarrow&  \pi^{-} N, \ \ \ \ \  \ \ \ \ \ \  \tau^{-} \rightarrow l^{-} \bar{\nu}_{l} N,
\end{eqnarray}
where $M = K, D, B$ and $l, l_{i}=e, \mu$. In the first decay of Eq. (\ref{DECAYS}) both $\nu$ denote the standard neutrino or antineutrino dominated by any of the neutrino flavors $\nu_{e}, \nu_{\mu}, \nu_{\tau}$.  These reactions include lepton number and flavor conserving as well as LFV and LNV decays. In the first case they receive the SM contributions, which alone give good agreement with the experimental data. 

\begin{figure}[htbp]
\centering
\includegraphics[width=0.8 \textwidth,bb=100 660 580 800]{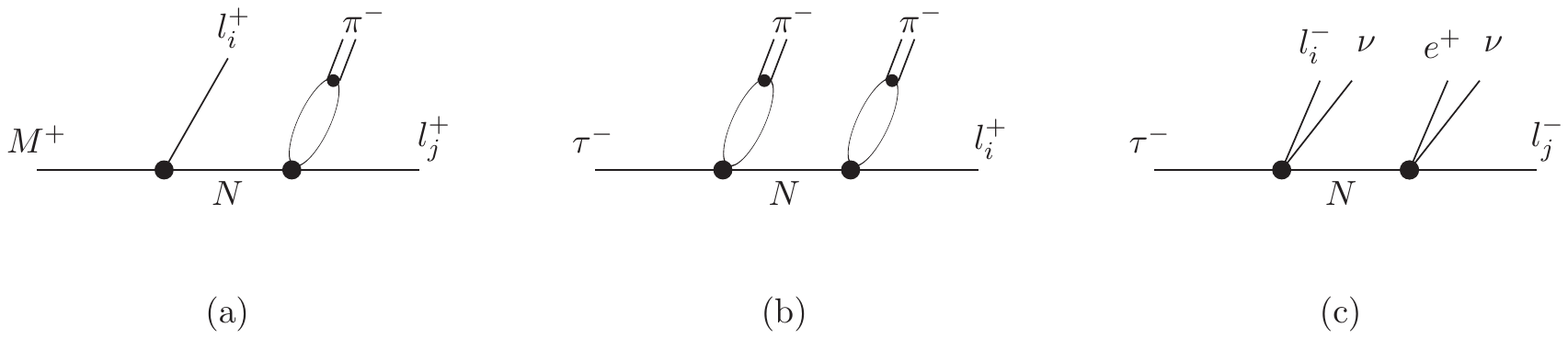}
\caption{Structure of the lowest order contribution of sterile neutrino N  to the semileptonic and leptonic meson and  $\tau$ 
decays. Here  $l_i, l_j = e, \mu$. } 
\label{fig-1} 
\end{figure}

The LFV and LNV decays (\ref{DECAYS}) are only possible beyond the SM. In the present framework they proceed  
according to the diagrams shown in Fig.\ref{fig-1} with 
sterile neutrino $N$ as a virtual particle.
Considering LNV decays we assume that sterile neutrino  is a Majorana particle $N=N^{c}$.
When the intermediate  sterile neutrino $N$ in these diagrams is off-shell their contribution to the processes (\ref{DECAYS}) is negligibly  small  \cite{tau-decPaper}, being far away from experimental reach. On the other hand there exist specific domains of sterile neutrino mass $m_{N}$ where N comes, for kinematical reasons, close to its mass-shell 
% can be considered with a good approximation \cite{tau-decPaper}  as a nearly on-shell intermediate state in diagrams Fig.  \ref{fig-1},
leading to 
%enormous 
resonant enhancement  \cite{K-decPaper, tau-decPaper, M-decPaper} of the diagrams in Fig.\ref{fig-1}.  These domains of $m_{N}$ will be specified below.

The decay rate formulas for the reactions in Eq.  (\ref{DECAYS}) can be directly calculated from the diagrams in Fig.\ref{fig-1} and Lagrangian (\ref{CC-NC}) for arbitrary mass $m_{N}$ of sterile neutrino.  We focus on the regions of $m_{N}$
where the sterile neutrino contribution is resonantly enhanced \cite{K-decPaper,tau-decPaper,M-decPaper}.   In these mass domains the intermediate sterile neutrino in Fig. \ref{fig-1} can be treated as nearly on-mass-shell state. This is to say, the sterile neutrino  $N$ is produced in the left vertices of the diagrams in Fig.\ref{fig-1}, propagate as a free unstable particle and then finally decays in the right vertices.  Thus the decay rate formulas for the reactions  $\tau, M \rightarrow X_{1}  X_{2} l$ can be represented in the form of products of the two factors:  $\tau$ or meson decay rate to the sterile neutrino 
$\Gamma (\tau, M \rightarrow N X_{1})$ and a branching ratio of  the sterile neutrino decay 
$Br(N\rightarrow l X_{2})$, where $X_{i}, l$ represent  final state particles of (\ref{DECAYS}). This representation is approximate and valid in the ``narrow width approximation'' $\Gamma_{N}\ll m_{N}$, where $\Gamma_{N}$ is the total decay width of sterile neutrino. As seen from Fig.\ref{fig-6},  this condition is satisfied in the region of $m_{N}$ studied in our analysis 
where $\Gamma_{N} < 10^{-10}$ MeV .  Below we list the decay rate formulas in this approximation for the reactions in Eq. (\ref{DECAYS}) specifying the corresponding 
resonant regions of $m_{N}$ where they are applicable. 
%For their derivation we refer to \cite{K-decPaper,tau-decPaper,M-decPaper}.
% and the corresponding decay rates 
These formulas are readily derived from the diagrams in Fig.\ref{fig-1}, considering the two vertices as the two independent processes of sterile neutrino production and its subsequent decay. 
%Then the only what is needed is to properly combine   

For semileptonic decays of mesons $M$ and $\tau$-lepton the decay rate formulas are
\begin{eqnarray}\label{DECRATEM}
%
%\Gamma^{ }(M^{+}\rightarrow \pi^{-} l^+ l^+)&\approx&  \Gamma(M^{+}\rightarrow l^+ N)\frac{\Gamma(N^{c}\rightarrow l^+ 
%\pi^{-})}{\Gamma_{N}}, \\
\Gamma^{ }(M^{+}\rightarrow \pi^{-} e^+ e^+)&\approx&  \Gamma(M^{+}\rightarrow l^+ N)\frac{\Gamma(N^{c}\rightarrow e^+ \pi^{-})}{\Gamma_{N}}, \\
 \label{DEC-LNV-MUE}
\Gamma^{ }(M^{+}\rightarrow \pi^{-} \mu^+ e^+) &\approx&
 \Gamma(M^{+}\rightarrow e^+ N)\frac{\Gamma(N^{c}\rightarrow \mu^+ \pi^{-})}{\Gamma_{N}} + \Gamma(M^{+}\rightarrow \mu^+ N)\frac{\Gamma(N^{c}\rightarrow e^+ \pi^{-})}{\Gamma_{N}},\\
 \label{DEC-LFV}
 \Gamma^{ }(M^{+}\rightarrow \pi^{+} \mu^{-} e^+)&\approx& 
 \Gamma(M^{+}\rightarrow e^+ N)\frac{\Gamma(N\rightarrow \mu^- \pi^{+})}{\Gamma_{N}},
 \end{eqnarray}
 valid in $m_e + m_\pi < m_N < m_M - m_{e}$,
 \begin{eqnarray}\label{DECRATE1}
\Gamma^{ }(\tau^{-}\rightarrow \pi^{-} \pi^{\pm} l^{\mp}) \approx
\Gamma(\tau^{-}\rightarrow \pi^{-}  N)\times \left\{\frac{\Gamma( N \rightarrow l^{-} \pi^{+})}{\Gamma_{N}},\ 
\frac{\Gamma( N^{c} \rightarrow l^{+} \pi^{-})}{\Gamma_{N}}
\right\}
\end{eqnarray}
valid in $m_l + m_\pi < m_N < m_\tau - m_\pi$.
Studying in subsection \ref{Purely leptonic decays} purely leptonic $\tau$-decays shown in Eq. (\ref{DECAYS}), we will need the decay rates summed over all the standard light neutrino and antineutrino in the final state. The corresponding formulas take the form
\begin{eqnarray}\label{DECRATE2}
 \Gamma^{ }(\tau^{-}\rightarrow e^{-} e^{+} e^{-} \nu  \nu) &\approx&  (1+ \delta_{N}) \sum_l \  \left[\Gamma(\tau^{-}\rightarrow e^{-}  \bar{\nu}_e N)\frac{\Gamma( N \rightarrow e^+ e^- \nu_l)}{\Gamma_{N}} + \right.\\
 \nonumber
&& \left. \ \ \ \ \ \ \ \ \ \ \ \ \  \ \ \  \ + \Gamma(\tau^{-}\rightarrow e^{-}  \nu_\tau N^{c})\frac{\Gamma( N^{c} \rightarrow e^+ e^- \bar{\nu}_l)}{\Gamma_{N}}\right],  \\[3mm]
 \Gamma^{ }(\tau^{-}\rightarrow e^{-} e^{+} \mu^{-} \nu  \nu)&\approx&  
 \Gamma(\tau^{-}\rightarrow e^{-} \bar{\nu}_e N)\frac{\Gamma(N\rightarrow e^+ \mu^- \nu_e)}{\Gamma_{N}} 
 + 
\delta_{N} \cdot \Gamma(\tau^{-}\rightarrow e^{-}  \bar{\nu}_e N)\frac{\Gamma( N^{c} \rightarrow e^+ \mu^-  \bar{\nu}_\mu)}{\Gamma_{N}} + 
 \label{teta2} \  \ \\ \nn 
&+& \delta_{N} \cdot \Gamma(\tau^{-}\rightarrow e^{-}  \nu_\tau N^{c})\frac{\Gamma( N \rightarrow e^+ \mu^- \nu_e)}{\Gamma_{N}} + \Gamma(\tau^{-}\rightarrow e^{-}  \nu_\tau N^{c})\frac{\Gamma( N^{c} \rightarrow e^+ \mu^-  \bar{\nu}_\mu)}{\Gamma_{N}}  + \\ \nn 
&+& (1+ \delta_{N}) \sum_l   \left[ \Gamma(\tau^{-}\rightarrow \mu^{-}  \bar{\nu}_\mu N)\frac{\Gamma( N \rightarrow e^+ e^- \nu_l)}{\Gamma_{N}} + \Gamma(\tau^{-}\rightarrow \mu^{-}  \nu_\tau N^{c})\frac{\Gamma( N^{c} \rightarrow e^+ e^- \bar{\nu}_l)}{\Gamma_{N}} \right] 
\end{eqnarray}
valid in $2 m_{e}< m_N < m_{\tau} - m_e$.  Here $\delta_{M} = 0,1$ for Dirac and Majorana case of sterile neutrino $N$, respectively. 
Summation in (\ref{DECRATE2}) and (\ref{teta2}) runs over $l=e, \mu, \tau$.
The partial decay rates $\Gamma(\tau, M\rightarrow X N)$ and  $\Gamma(N\rightarrow Y l)$  and the total decay rate of sterile neutrino $\Gamma_{N}$ involved in Eqs. (\ref{DECRATEM})-(\ref{teta2}) are specified in  Appendix.  
Implicitly all the partial decay rates include the corresponding threshold step-functions. 
For further convenience we rewrite Eq.(\ref{DECRATE2}), (\ref{teta2}) in the form
\begin{eqnarray}\label{DECRATES}
\Gamma^{ }(\tau^{-}\rightarrow e^{-} e^{+} e^{-} \nu  \nu)&\approx&  (1 + \delta_N) \frac{ \ \Gamma_\tau^{(e \nu N) } \Gamma_N^{(e e \nu_\tau)} }{\Gamma_N}(|U_{\tau N}|^4 +  |U_{\mu N}|^2 |U_{\tau N}|^2+ (\beta + 1) |U_{e N}|^2 |U_{\tau N}|^2 + \\ \nonumber 
&+&|U_{e N}|^2 |U_{\mu N}|^2 + \beta |U_{e N}|^4), \  \   \\[3mm]
 \label{DECRATES2} 
 \Gamma^{ }(\tau^{-}\rightarrow e^{-} e^{+} \mu^{-} \nu  \nu) &\approx&   \frac{ \ \Gamma_\tau^{(e \nu N) } \Gamma_N^{(e \mu \nu )} \ + \ \Gamma_\tau^{(\mu \nu N) } \Gamma_N^{(e e \nu_\tau)}}{\Gamma_N}
( (1 + \delta_N) \alpha_2 |U_{\tau N}|^4 + (\alpha_1 + 2 (1 + \delta_N) \alpha_2) |U_{\mu N}|^2 |U_{\tau N}|^2+\ \ \ \ \ \  \\ \nonumber 
&+&(\delta_N \alpha_1 + (1 + \delta_N) \beta \alpha_2) |U_{e N}|^2 |U_{\tau N}|^2 + (\delta_N \alpha_1 + (1 + \delta_N) \beta \alpha_2) |U_{e N}|^2 |U_{\mu N}|^2 + \alpha_1 |U_{e N}|^4 + \\ \nn &+& (1 + \delta_N) \alpha_2 |U_{\mu N}|^4), 
\end{eqnarray}
where 
\begin{eqnarray}\label{beta}
\beta &=& \Gamma_N^{(ee\nu_e)}/\Gamma_N^{(ee\nu_\tau)} \approx 4.65,\\
\label{ALPHA}
\alpha_1 &=& \frac{\Gamma_\tau^{(e\nu N)} \Gamma_N^{(e\mu \nu)}}{\Gamma_\tau^{(e \nu N) } \Gamma_N^{(e \mu \nu )} + \ \Gamma_\tau^{(\mu \nu N) } \Gamma_N^{(e e \nu_\tau)} }, \ \
\alpha_2 = \frac{\Gamma_\tau^{(\mu \nu N)} \Gamma_N^{(e e \nu_\tau)} }{\Gamma_\tau^{(e \nu N) } \Gamma_N^{(e \mu \nu )} + \ \Gamma_\tau^{(\mu \nu N) } \Gamma_N^{(e e \nu_\tau)}}.
\end{eqnarray}
In Eqs. (\ref{DECRATES})-(\ref{ALPHA}) we used notations $\Gamma_N^{(ll\nu)}, \Gamma_\tau^{(l \nu N)}$ introduced in Eqs. (\ref{DECAYRATETAU2})-(\ref{nuqq}).

As we already mentioned, in the resonant regions of the sterile neutrino mass $m_{N}$, specified in 
Eqs. (\ref{DECRATEM})-(\ref{teta2}), the intermediate  sterile neutrino  $N$, produced in $\tau$ and meson $M$ decays 
(see Fig.\ref{fig-1}), propagates as a real particle and decays at certain distance from the production point. If this distance is larger than the size of the detector, the sterile  neutrino escapes from it before decaying and the signature of 
$\tau\rightarrow l\pi\pi$, $\tau\rightarrow  e e l \nu \nu$ or $M \rightarrow \pi  l l$ cannot  be recognized. In this case in order to calculate the rate of $\tau$ or meson decay within a detector one should multiply the theoretical expressions $\Gamma^{ }$ in (\ref{DECRATEM})-(\ref{teta2}) by the probability  $P_{N}$ of sterile neutrino decay within a detector of the size $L_{D}$. Within reasonable approximations it takes the form 
\cite{Atre:2009rg}
\begin{eqnarray}\label{PN}
P_{N}\approx1-exp(-L_{D}\Gamma_{N}),
\end{eqnarray}
where $\Gamma_{N}$ is the total decay rate of sterile neutrino calculated in (\ref{GT-comp-incl}).

\begin{figure}[htbp]
\centering
\includegraphics[width=0.6\textwidth]{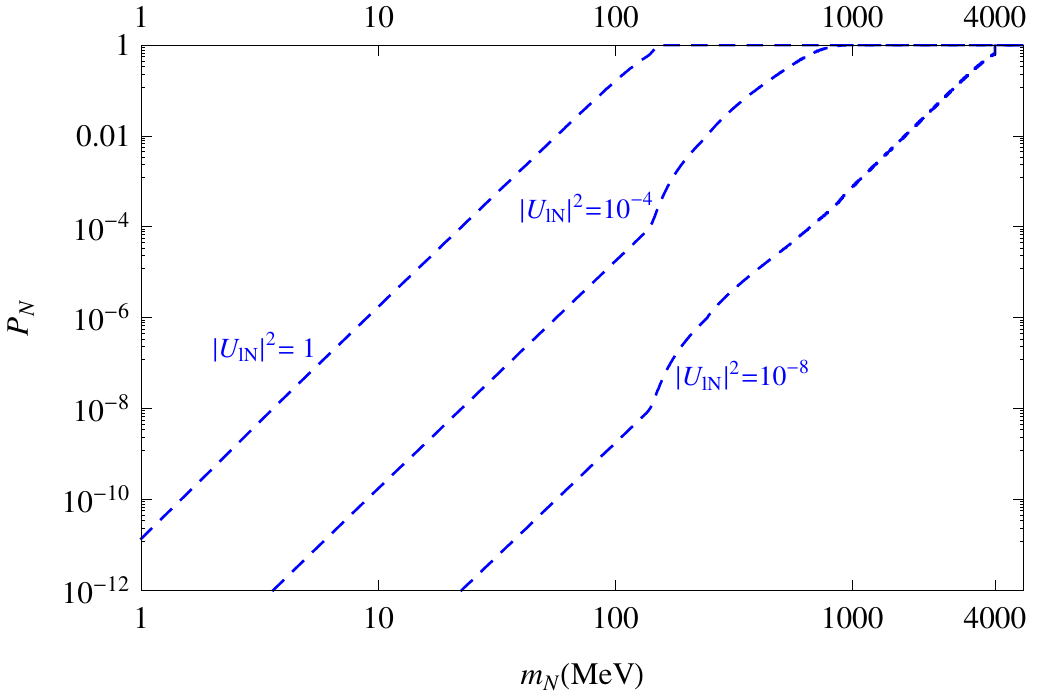}
\caption{The probability  $P_{N}$ of sterile neutrino decay within a detector of the size of $L_{D}=10$ m versus sterile neutrino mass $m_{N}$ for several values of mixing matrix elements $|U_{l N}|^{2}$, 
assuming $|U_{e N}|^{2}=|U_{\mu N}|^{2}=|U_{\tau N}|^{2}$.} 
\label{fig-2}
\end{figure}

Then,  the rates $\Gamma^{D}$ of $\tau$ and meson decays within detector volume should be estimated according to
\begin{eqnarray}\label{DR-PN}
\Gamma^{D} = \Gamma \times P_{N} ,
\end{eqnarray}
where $\Gamma$ are decay rates given by Eqs. (\ref{DECRATEM})-(\ref{teta2}).
%This situation clearly depends on the the concrete  experimental setup, particularly on the size of the detector. 
In our numerical analysis we take for concreteness $L_D = 10m$ which is typical for this kind of experiments.  
In Fig. \ref{fig-2} we plotted $P_{N}$ v.s. sterile neutrino mass $m_{N}$ for several values of mixing matrix elements 
$|U_{lN}|^{2}$. For illustration of typical tendencies we assumed in this plot  $|U_{e N}|^{2}=|U_{\mu N}|^{2}=|U_{\tau N}|^{2}$.
We do not use this assumption in our analysis.
As seen, $P_{N}$ becomes small for $m_{N}<100$ MeV even for rather large values of $|U|^{2}_{lN}$.  Thus, in this region of 
$m_{N}$ the effect of finite size of detector,  described by $P_{N}$, significantly affects the decay rates of the studied processes and should be taken into account.

\section{Limits on Sterile Neutrino  Mixing $U_{\tau N}$}
\label{sec-3}

In the literature there are various limits on the mixing parameters $U_{\alpha N}$ (with $\alpha = e,\mu, \tau$)
extracted from direct and indirect experimental searches \cite{PDG} for this particle, in a wide 
region of its mass. A recent summary of these limits, extracted from the corresponding 
experimental data, can be found in Ref. \cite{Atre:2009rg}.  In the present paper we focus on the least constrained mixing parameter $U_{\tau N}$. 
In Fig. \ref{fig-3} we show the exclusion plots for $|U_{\tau N}|^{2}$ existing in the literature \cite{Atre:2009rg} together with our exclusion curves  derived in the present section. For derivation of these curves we will analyze sterile neutrino contribution  to the decays listed in (\ref{DECAYS})-(\ref{TAU-N-th}).
%contribution purely leptonic $\tau$-decays, joint analysis of leptonic and semileptonic decays of $\tau$ together with %semileptonic decays of  $K, D, B$ mesons. 

As seen from Eqs. (\ref{DECRATEM})-(\ref{teta2})  the decay rates of  the processes (\ref{DECAYS}) depend on all the three 
$U_{\alpha N}$ (with $\alpha = e,\mu, \tau$)  mixing matrix elements. In the literature it is common practice to adopt some ad hoc assumptions on their relative size in order to extract limits on them from the experimental bounds on the  corresponding  decay rates. In particular, limits from CHARM  \cite{CHARM} and NOMAD \cite{NOMAD} plotted in Fig.  \ref{fig-3} assume 
$|U_{\tau N}| \gg |U_{e N}|, |U_{\mu N}|$.
These  assumptions may reduce reliability of the obtained limits.   
Below we derive analytic expressions for  limits on $|U_{\tau N}|^2$ in different mass ranges of $m_N$ without any kind of such assumptions.

\subsection{Purely leptonic decays}
\label{Purely leptonic decays}

First we exploit for extraction of $|U_{\tau N}|$ the following experimental results for the branching ratios of purely leptonic $\tau$-decays
%In order to constrain $U_{\tau N}$ in this mass domain we use the experimental data for the purely leptonic decays 
%
\cite{PDG}
\begin{eqnarray}\label{Data-PL}
Br(\tau^{-}\rightarrow e^{-} e^{+} e^{-} \bar{\nu}_{e}  \nu_{\tau}) &=& (2.8 \pm 1.5) \times 10^{-5} \\
\label{Data-PL-1} 
Br(\tau^{-}\rightarrow e^{-} e^{+} \mu^{-} \bar{\nu}_{\mu}  \nu_{\tau}) &<& 3.6 \times 10^{-5}.
\end{eqnarray}
The first decay has been observed experimentally and its experimentally measured branching ratio agrees with the SM prediction within the standard deviation $\Delta^{exp}(\tau^{-}\rightarrow e^{-} e^{+} e^{-} \nu  \nu) =1.5\times 10^{-5}$.
Neutrino assignment in the final states of the decays (\ref{Data-PL})-(\ref{Data-PL-1}) corresponds to what is suggested by the SM. However, in the experiments, measuring these decays, the final state neutrinos cannot be actually identified. Therefore, considering beyond the SM mechanisms with LFV one should take into account the possibility that all the light neutrinos 
$\nu_{e}$, $\nu_{\mu}$ and $\nu_{\tau}$ may contribute to the final state of  the decays (\ref{Data-PL})-(\ref{Data-PL-1}).  
Formulas (\ref{DECRATES})-(\ref{DECRATES2})  were derived for the very this case. They describe the sterile neutrino resonant contribution (diagram 
Fig.\ref{fig-1}(a)) to the decays (\ref{Data-PL})-(\ref{Data-PL-1}) and will be used in the analysis of this subsection.

We also assume that the sterile neutrino contribution to the process (\ref{Data-PL}), if exists, should be less than $\Delta^{exp}$.
For the decay (\ref{Data-PL-1}), not yet observed experimentally, there exists only the above indicated upper bound and the sterile neutrino contribution has to obey this bound. 
%Below we derive limits on $|U_{\tau N}|$ from the decay   (\ref{Data-PL}). The corresponding derivation and resulting limits on %$|U_{\tau N}|$ from the experimental bound  
%(\ref{Data-PL-1}) are very similar and not presented here for brevity.

Taking into account the finite detector size effect according to Eq. (\ref{DR-PN}) we write for decay rate $\Gamma^{D}$ within detector volume  
\begin{eqnarray}\label{PUreLep1}
\Gamma^{D}(\tau^{-}\rightarrow e^{-} e^{+} l^{-} \nu  \nu) \approx \Gamma(\tau^{-}\rightarrow e^{-} e^{+} l^{-} \nu  \nu )\times P_{N},
\end{eqnarray}
with $\Gamma(\tau^{-}\rightarrow e^{-} e^{+} l^{-} \nu  \nu )$ given by  (\ref{DECRATE2}), (\ref{teta2}). 
As we discussed in the previous section, the probability $P_N$ of sterile neutrino  decay within detector becomes rather small for $m_N < 100$ MeV.  Therefore, in this mass range we may approximate the expression in (\ref{PN}) by 
$P_{N}\approx L_{D} \Gamma_{N}$.  This is a reasonable approximation for this part of our analysis since the limits, which will be obtained here, correspond to the exclusion curve (a) in Figs. \ref{fig-3} and curves in 
\mbox{Figs. \ref{fig-4}, \ref{fig-5}} located in the region 
$m_{N}\leq 100$ MeV, where $L_{D} \Gamma_{N} \sim 0.01$. 

In this approximation we find from (\ref{DECRATES}) and  (\ref{PUreLep1})
%For the more simple case of $\tau^{-}\rightarrow e^{-} e^{+} e^{-} \nu  \nu$ in this approximation we obtain
\begin{eqnarray}\label{LIMIT1} 
\Gamma^{D}(\tau^{-}\rightarrow e^{-} e^{+} e^{-} \nu  \nu) &\approx&  (1+\delta_N) \Gamma_\tau^{(e \nu N) } \Gamma_N^{(e e \nu_\tau)} L_D   (  |U_{\tau N}|^4 +  |U_{\mu N}|^2 |U_{\tau N}|^2+ (\beta + 1) |U_{e N}|^2 |U_{\tau N}|^2 +  \\ \nn + \  |U_{e N}|^2 |U_{\mu N}|^2  &+& \beta |U_{e N}|^4). 
\end{eqnarray}
According to our assumption, discussed after Eqs. (\ref{Data-PL})-(\ref{Data-PL-1}), we require 
\begin{eqnarray}\label{Delta-1}
\tau_{\tau }\Gamma^{ }(\tau^{-}\rightarrow e^{-} e^{+} e^{-} \nu  \nu)\leq \Delta^{exp}(\tau^{-}\rightarrow e^{-} e^{+} e^{-} \nu  \nu) \approx 1.5\times 10^{-5}, 
\end{eqnarray}
where 
$\tau_{\tau} = (290.6\pm 1.0) \times 10^{-15}$ s  is the  $\tau$-lepton mean life \cite{PDG}.  Then we obtain the following upper limits
\begin{eqnarray}\label{LIMIT3}
|U_{\tau N}|^2 \leq \sqrt \frac{\Delta^{exp}(\tau^{-}\rightarrow e^{-} e^{+} e^{-} \nu  \nu)}{ \ \Gamma_\tau^{(e \nu N) } \Gamma_N^{(e e \nu_\tau)} (1 + \delta_N) L_D\ \tau_{\tau} } ,
\end{eqnarray}
\begin{eqnarray}\label{LIMIT32}
|U_{\tau N} U_{\mu N}| \leq \sqrt \frac{\Delta^{exp}(\tau^{-}\rightarrow e^{-} e^{+} e^{-} \nu  \nu)}{ \ \Gamma_\tau^{e \nu N } \Gamma_N^{e e \nu_\tau} (1 + \delta_N) L_D\  \tau_{\tau}} \  , \ \ \ 
|U_{\tau N} U_{e N}| \leq \sqrt \frac{\Delta^{exp}(\tau^{-}\rightarrow e^{-} e^{+} e^{-} \nu  \nu)}{ ( \beta + 1) \ \Gamma_\tau^{e \nu N } \Gamma_N^{e e \nu_\tau} (1 + \delta_N) L_D\ \tau_{\tau} } \ .
\end{eqnarray}
Similarly, we derive limits based on the experimental bound (\ref{Data-PL-1}). Using Eq. (\ref{DECRATES2}), we find   
\begin{eqnarray}\label{LIMIT4}
|U_{\tau N}|^2 &\leq& \sqrt \frac{Br^{exp}(\tau^{-}\rightarrow \mu^{-} e^{+} e^{-} \nu  \nu)}{ ( \ \Gamma_\tau^{(e \nu N) } \Gamma_N^{(e \mu \nu)} +  \Gamma_\tau^{(\mu \nu N) } \Gamma_N^{(e e \nu_\tau)}\ ) (1 + \delta_N) \alpha_2 L_D \ \tau_{{\tau}}}, \\
\label{LIMIT42}
|U_{\tau N} U_{\mu N}|  &\leq& \sqrt \frac{Br^{exp}(\tau^{-}\rightarrow \mu^{-} e^{+} e^{-} \nu  \nu)}{ ( \ \Gamma_\tau^{(e \nu N) } \Gamma_N^{(e \mu \nu)} +  \Gamma_\tau^{(\mu \nu N) } \Gamma_N^{(e e \nu_\tau)} \ )(\alpha_1 + 2 (1 + \delta_N) \alpha_2 ) L_D\ \tau_{\tau} }, \\
\label{LIMIT43}
|U_{\tau N} U_{e N}| &\leq&  \sqrt \frac{Br^{exp}(\tau^{-}\rightarrow \mu^{-} e^{+} e^{-} \nu  \nu)}{ ( \ \Gamma_\tau^{(e \nu N) } \Gamma_N^{(e \mu \nu)} +  \Gamma_\tau^{(\mu \nu N) } \Gamma_N^{(e e \nu_\tau)} \ )
(\delta_N \alpha_1 + (1 + \delta_N) \beta \alpha_2 ) L_D\ \tau_{\tau} }. \ \ \ 
\end{eqnarray}
Here, $Br^{exp}$ denotes left-hand side of the experimental bound in (\ref{Data-PL-1}). The limits 
(\ref{LIMIT3})-(\ref{LIMIT43}) are plotted in Fig. \ref{fig-3}-\ref{fig-5} for the case of sterile Majorana neutrino. Drawing the exclusion curves, we selected the most stringent limit among (\ref{LIMIT3})-(\ref{LIMIT43}) for each mass value $m_{N}$ within the studied mass range. As seen, the present experimental data 
(\ref{Data-PL})-(\ref{Data-PL-1}) on purely leptonic $\tau$-decays set rather weak constraints on $|U_{\tau N}|$ and on 
$|U_{\tau N} U_{e N}|$, $|U_{\tau N} U_{\mu N}|$  in the mass region 
$1\mbox{MeV}\leq m_{N} \leq 100$ MeV. Our limits on $|U_{\tau N}|$, corresponding to the curve (a) in Fig \ref{fig-3}, are significantly weaker than the limitations from other searches shown in Fig. \ref{fig-3}.  However, our limits for
$|U_{\tau N} U_{e N}|$ and $|U_{\tau N} U_{\mu N}|$  in Figs. \ref{fig-4}, \ref{fig-5} to our best knowledge are new in this mass region.  

\subsection{Leptonic and semileptonic decays}

Now we combine the purely leptonic $\tau$-decays considered in the previous subsection with the semileptonic decays of
$\tau$ and $K, D$-mesons using the experimental data (\ref{Data-PL}), (\ref{Data-PL-1}) and the experimental limits on the following branching ratios \cite{PDG}:
\begin{eqnarray}\label{DATA-TAU-1}
Br(\tau^{-}\rightarrow \pi^{-} \pi^{+} e^{-}) &\leq& 1.2 \times 10^{-7},\ \ \ \ 
Br(\tau^{-}\rightarrow \pi^{-} \pi^{+} \mu^{+}) \leq 7 \times 10^{-8},\\ 
\label{DATA-K-E}
Br(K^{+}\rightarrow \pi^{-} e^{+} e^{+}) &\leq& 6.4 \times 10^{-10} ,\ \ \
%\label{DATA-K-MU}
Br(K^{+}\rightarrow \pi^{+} \mu^{-} e^{+}) \leq 1.3 \times 10^{-11},\\
\label{DATA-D-MU}
Br(D^{+}\rightarrow \pi^{-} e^{+} e^{+}) &\leq&  3.6 \times 10^{-6},\ \ \ \
%\label{DATA-D-MU} 
Br(D^{+}\rightarrow \pi^{+} \mu^{-} e^{+}) \leq  3.4 \times 10^{-5}.
\end{eqnarray}
Assuming that in all these decays sterile neutrino N contributes resonantly we should limit ourselves to the mass domain:
\begin{eqnarray}\label{Mass-dom-1}
%245\  \mbox{MeV} \leq m_N \leq m_{\tau} - m_{\pi} \approx 1637\  \mbox{MeV}.
m_{\pi} + m_{\mu} \approx 245\  \mbox{MeV} \leq m_N \leq m_{\tau} - m_{\pi} \approx 1637\  \mbox{MeV}.
\end{eqnarray}
Within this mass domain the experimental bounds  (\ref{DATA-K-E}) contribute to our analysis only up to
\mbox{$m_{N}\leq m_{K}-m_{e} \approx 493.2$ MeV} corresponding to the mass range of the resonant contribution of sterile neutrino to these decays of K-meson. In the above list (\ref{DATA-TAU-1})-(\ref{DATA-D-MU})  one could also include the existing experimental bounds on the other LNV and LFV decays of $\tau$ and $D_{s}, B$ mesons. However, they have negligible impact on our results presented below.

In this part of our analysis we put $P_{N}=1$ for the probability (see Eq.  (\ref{PN})) of decay of nearly on-mass-shell sterile neutrino,  resonantly contributing to the analyzed processes.  Thus we assume that these processes occur completely within a detector volume. This is a good approximation for the case of the limits on $U_{\tau N}$, which will be derived here and displayed in Fig. \ref{fig-3} as curve (b).  To see this one can check the plot for $P_{N}$ shown in Fig. \ref{fig-2}.
 
In the mass domain (\ref{Mass-dom-1}) we can use Eqs. (\ref{DECRATEM})-(\ref{DECRATES2}) for the corresponding decay rates. Below we combine these formulas in a system of equations. Solving them with respect to $|U_{\tau N}|$ and applying the experimental bounds (\ref{Data-PL}), (\ref{Data-PL-1}) and  (\ref{DATA-TAU-1})-(\ref{DATA-D-MU}) we find upper limits on this mixing parameter.  
For our purpose it is sufficient to use ether of the two experimental bounds (\ref{Data-PL}), (\ref{Data-PL-1}). 
%We use the first one. 
%For this procedure we need only one of the two bounds (\ref{Data-PL}), (\ref{Data-PL-1}). 
We select  (\ref{Data-PL}) which leads to a bit more stringent limits on $|U_{\tau N}|$.  

Let us introduce the following notations
\begin{eqnarray}\label{F}
F_{ee}(\tau) &=& \frac{\Delta^{exp}(\tau^-\rightarrow e^- e^+e^- \nu \nu) }{ \ (1 + \delta_N) \ \Gamma_\tau^{(l \nu N)} 
\Gamma_N^{(e e \nu_\tau)}\ \tau_{\tau}},  \ \
%F_{e\mu}(\tau) = \frac{Br^{exp}(\tau^-\rightarrow e^- e^+\mu^- \nu \nu) }{ \ \Gamma_\tau^{(l \nu N)} 
%\Gamma_N^{(e e \nu_\tau)}\ \tau_{\tau}}, \\ \nonumber
F_{\pi l}(\tau) = \frac{Br^{exp}(\tau^-\rightarrow \pi^-\pi^\pm l^\mp) }{\Gamma_\tau^{(\pi N)} \Gamma_N^{(l \pi)} 
\ \tau_{\tau}} , \\ \nn
F_{ee}(M) &=& \frac{Br^{exp}(M^+ \rightarrow \pi^- e^+e^+ ) }{ \ \Gamma_M^{(e N)} \Gamma_N^{(e \pi)} \ \tau_{M}}, \ \  \  \ \ 
F_{e \mu}(M) = \frac{Br^{exp}(M^+ \rightarrow \pi^- \mu^+e^+ ) }{ \ ( \Gamma_M^{(e N)} \Gamma_N^{(\mu \pi)} + \Gamma_M^{(\mu N)} \Gamma_N^{(e \pi)}) \ \tau_{M}} ,
\end{eqnarray}
where $\tau_{\tau}, \tau_{M}$ are mean lives of $\tau$ and $M=K^{+}, D^{+}$;  the right-hand sides of the experimental bounds in   (\ref{DATA-TAU-1})-(\ref{DATA-D-MU}) are denoted by $Br^{exp}$; the quantity 
$\Delta^{exp}$ was introduced after Eqs. (\ref{Data-PL}) and (\ref{Data-PL-1}). 

Now we can rewrite the experimental limits on (\ref{Data-PL}) and  (\ref{DATA-TAU-1})-(\ref{DATA-D-MU}) in the form
\begin{eqnarray}\label{F1}
\frac{ |U_{\tau N}|^4 +  |U_{\mu N}|^2 |U_{\tau N}|^2+ (\beta + 1) |U_{e N}|^2 |U_{\tau N}|^2 + |U_{e N}|^2 |U_{\mu N}|^2 + \beta |U_{e N}|^4)}{a_{e}  |U_{eN}|^{2}  + a_{\mu}   |U_{\mu N}|^{2} +  
a_{\tau}   |U_{\tau N}|^{2}} \leq F_{ee}(\tau), \\ 
\frac{ |U_{\tau N}|^2 |U_{l N}|^2 }{a_{e}  |U_{eN}|^{2}  + a_{\mu}   |U_{\mu N}|^{2} +  
a_{\tau}   |U_{\tau N}|^{2}} \leq F_{\pi l}(\tau), \label{F2} \\
\frac{ |U_{e N}|^2 |U_{l N}|^2 }{a_{e}  |U_{eN}|^{2}  + a_{\mu}   |U_{\mu N}|^{2} +  
a_{\tau}   |U_{\tau N}|^{2}} \leq F_{e l}(M). \label{F3}
\end{eqnarray}
Here $l = e, \mu$. Solving (\ref{F1})-(\ref{F3}) we find
\begin{eqnarray}\label{UT2}
|U_{\tau N}|^2 \leq c_1 F_{ee}(\tau) + c_2 F_{\pi e}(\tau)+ c_3 F_{\pi \mu}(\tau)+ c_4 F_{ee}(M)+ c_5 F_{e \mu}(M).
\end{eqnarray}
%
%\ba{UT}
%|U_{\tau N}|^2 + (\beta - 1) |U_{e N}|^2 \leq c_1 F_{ee}(\tau) + c_2 F_{\pi e}(\tau)+ c_3 F_{\pi \mu}(\tau)+ c_4 F_{ee}(M)+ c_5 %F_{e \mu}(M),
%\ea
where 
\begin{eqnarray}\label{C}
c_1 = a_\tau, \  c_2 = a_e - 2 a_\tau , \  c_3 = a_\mu - a_\tau, \  c_4 =  (\beta-1)  a_e - \beta a_\tau, \ c_5 =  (\beta-1) a_\mu - a_\tau . 
\end{eqnarray}
We have checked that in the mass region (\ref{Mass-dom-1}) all the coefficients $c_{i}>0$.  The parameter $\beta$ is defined in  
(\ref{beta}). 
%Also $\beta-1>0$ as follows from the definition (\ref{beta}).   
%Taking these facts into account we finally find the desired limit
%\begin{eqnarray}\label{UT2}
%|U_{\tau N}|^2 \leq c_1 F_{ee}(\tau) + c_2 F_{\pi e}(\tau)+ c_3 F_{\pi \mu}(\tau)+ c_4 F_{ee}(M)+ c_5 F_{e \mu}(M).
%\end{eqnarray}
%
We plotted the corresponding exclusion curve in Fig.\ref{fig-3} labeled by (b) for the case of Majorana sterile neutrino. As seen, our limits are more stringent than the existing ones from CHARM \cite{CHARM} and DELPHI \cite{DELPHI} experiments in the sterile neutrino mass region
\mbox{300 MeV$\leq m_{N}\leq$900 MeV}. Note that in difference from the existing limits on $|U_{\tau N}|$  our limits are model independent in the sense that we have not made any assumptions on the other two mixing parameters $|U_{e N}|$ and $|U_{\mu N}|$. Instead, we excluded them combining the experimental limits on the branching ratios of different processes 
(\ref{Data-PL}), (\ref{Data-PL-1}) and  (\ref{DATA-TAU-1})-(\ref{DATA-D-MU}).

%%%%%%%%%%%%%%
\subsection{Sterile neutrino in the final state}

%There exist other processes observed experimentally 

Other experimental data which we apply for deriving limits on $U_{\tau N}$  are \cite{PDG}
\begin{eqnarray}
\label{N-fin-state-1}
Br(\tau^- \rightarrow l \bar \nu_l \nu_\tau) &=& (17.85 [17.36]\pm 0.05 )\%, \\
\label{N-fin-state-2}
Br(\tau^- \rightarrow \pi^-  \nu_\tau) &=& (10.91 \pm 0.07) \%,
\end{eqnarray}
where in the first line the central value 17.85 corresponds to $l=e$ and 17.36 to $l=\mu$. 
Both these experimental results agree with the SM predictions within the standard 
deviations $\Delta^{exp}(\tau\rightarrow l \nu \nu) = 0.05\%$ and $\Delta^{exp}(\tau \rightarrow l \pi \nu) = 0.07\%$.  
We already commented in subsection \ref{Purely leptonic decays} (after Eqs. (\ref{Data-PL}), (\ref{Data-PL-1})), that 
in the reported experimental results like in Eqs. 
(\ref{Data-PL})-(\ref{Data-PL-1}) and (\ref{N-fin-state-1})-(\ref{N-fin-state-2})  the final state neutrino assignment 
$\nu_{e,\mu,\tau}$ is made according to  what is suggested by the SM. However, in the experiments, measuring these decays, the final state neutrinos cannot be actually identified and are observed as a missing energy signature.
%observed events taken into account in (\ref{Data-PL})-(\ref{Data-PL-1}) have the signature of missing momentum 
Therefore, it is liable to imagine that instead of one or even both of the standard light neutrinos in the final states of decays in 
(\ref{N-fin-state-1})-(\ref{N-fin-state-2}) there may occur 
%in the final states of these decays may appear 
 some other neutral particles such as sterile neutrinos. We assume that in these modes of $\tau$-decay 
 appears one sterile neutrino $N$ accompanied by any of  $\nu_{e}, \nu_{\mu}, \nu_{\tau}$. Its mass must satisfy to 
$m_{N}\leq m_{\tau} - m_{l}$ and  
$m_{N}\leq m_{\tau} - m_{\pi}$ for the decays  (\ref{N-fin-state-1}) and (\ref{N-fin-state-2}) respectively.
 We also assume that this contribution, if exists, should be less than the corresponding standard deviation $\Delta^{exp}$.
%For the decay (\ref{Data-PL-1}), not yet observed experimentally, there is only the above indicated upper bound and the sterile %neutrino contribution has to obey this bound. %On the other hand we can apply to them the same reasoning, which  we discussed after Eqs. (\ref{Data-PL}), (\ref{Data-PL-1}), % and assume 
%On the other hand in both cases the final state neutrinos 
%manifest themselves only as a missing momentum in the vertices with $\pi$ or $e, \mu$. 
%They are not identified to be $\nu_{e}$ or $\nu_{\mu}$ as it is displayed in Eqs. (\ref{N-fin-state-1}), (\ref{N-fin-state-2}).
%This is just a hypothesis inspired by the SM prediction. Therefore, one may assume 
%Since (\ref{N-fin-state-1}), (\ref{N-fin-state-2}) are in agreement with the SM 

The contribution of  sterile neutrino $N$ to (\ref{N-fin-state-1}), (\ref{N-fin-state-2}) in the form
\begin{eqnarray}\label{St-Nu-fin-3}
\tau^- \rightarrow l \bar \nu_l N, \ \ \ \  \tau^- \rightarrow \pi^-  N
\end{eqnarray}
should be less than the corresponding $\Delta^{exp}$ since (\ref{N-fin-state-1}), (\ref{N-fin-state-2}) are in agreement with the SM. 
  
\begin{figure}[htbp]
\centering
\includegraphics[width=0.8\textwidth]{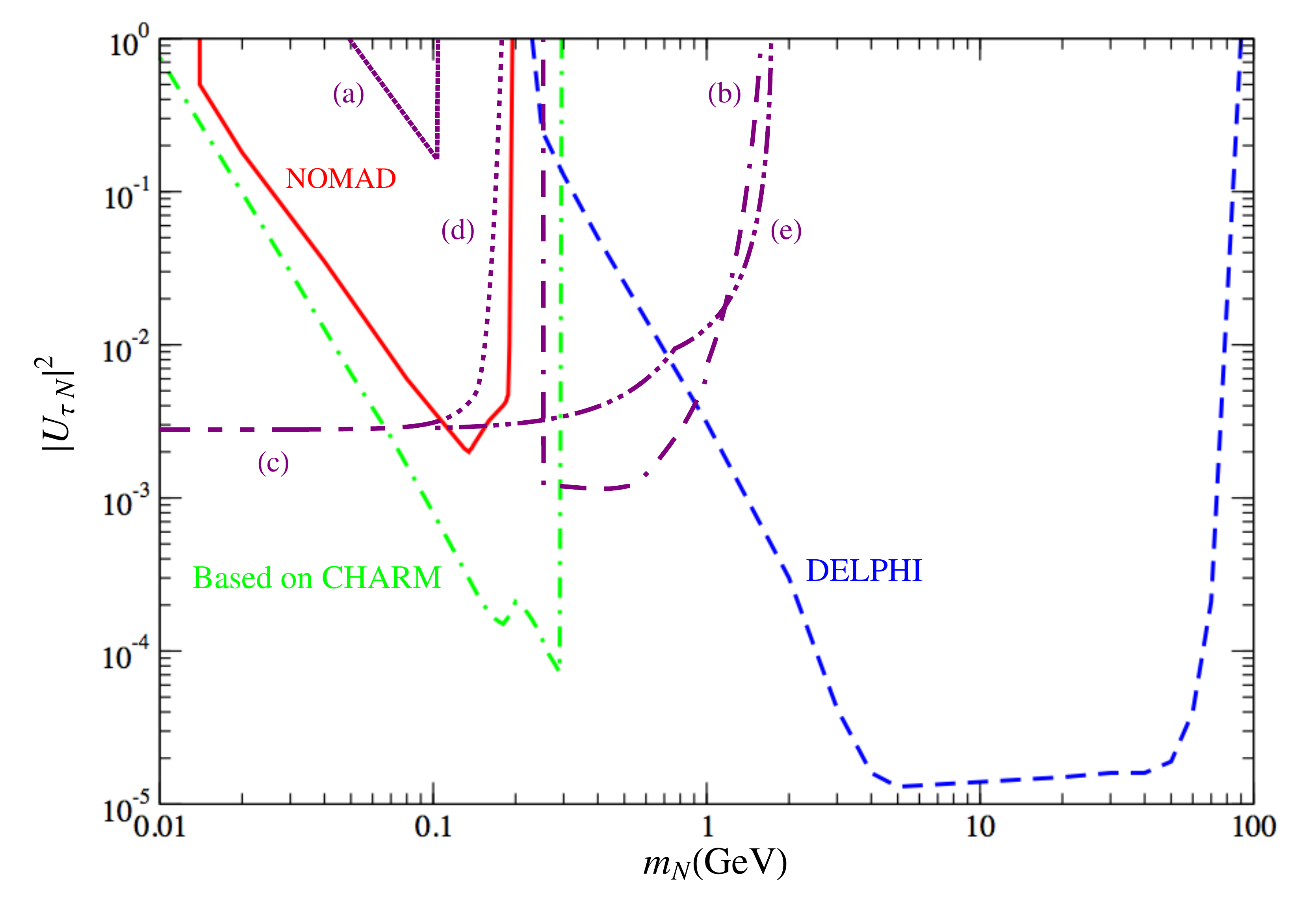}
\caption{Exclusion curves for $|U_{\tau N}|^{2}$ from the present analysis, denoted by (a)-(e), and the exclusion curves existing in the literature derived from  CHARM \cite{CHARM}, NOMAD \cite{NOMAD} and DELPHI \cite{DELPHI}
searches for sterile neutrino decays. The latter curves are taken from Ref. \cite{Atre:2009rg}. } 
\label{fig-3}
\end{figure}

Therefore, using (\ref{Dec-TAU-PI-N}) and  (\ref{DECAYRATETAU}) we find the limits 
 \begin{eqnarray}\label{LIMIT0}
|U_{\tau N}|^2 \leq 
\mbox{Min} \left\{\frac{\Delta^{exp}(\tau^{-}\rightarrow \pi^{-} \nu)}{ \ \Gamma_\tau^{(\pi N) }},  \frac{\Delta^{exp}(\tau^{-}\rightarrow \nu \nu l^{-} )}{ \ \Gamma_\tau^{( N \nu l) }}\right\},
\end{eqnarray}
where the minimal of the two values in the curl brackets are selected for each value of $m_{N}$. The corresponding exclusion curve is shown in Fig.\ref{fig-3} and comprises the two parts  (c) and (e). The part  (c) is dominated by the constraints on purely leptonic  $\tau$-decay mode while the part (e) is mainly due to the semileptonic mode  
shown in (\ref{St-Nu-fin-3}).  The exclusion curve (c), (e) cover a mass region 
\mbox{$0 \leq m_{N}\leq m_{\tau}-m_{\pi} \approx 1640$ MeV.}  This curve sets new limits on $U_{\tau N}$ for 0 $\leq m_{N}\leq 70$ MeV
and 300 MeV $\leq m_{N} \leq$ 700 MeV. In the region 500 MeV $\leq m_{N} \leq$ 700 MeV they are less stringent than our limits derived in the previous subsection from the data (\ref{Data-PL}), (\ref{Data-PL-1}),  (\ref{DATA-TAU-1})-(\ref{DATA-D-MU}) and corresponding to the curve (b) in Fig. \ref{fig-3}.  For $m_{N}\leq$ 100 MeV the part (c) of our exclusion curve is nearly constant 
and our limits for this mass range can be displayed as
\begin{eqnarray}\label{lim-100-MeV}
|U_{\tau N}|^{2} \leq 2.9 \times 10^{-3}, \ \ \ \ \ \mbox{for} \ \ \ \ \ 0\leq m_{N}\leq 100 \mbox{MeV}.
\end{eqnarray}

As we discussed previously, sterile neutrino produced in (\ref{St-Nu-fin-3})  can  decay within a detector with a probability $P_{N}$ defined  in (\ref{PN}). This would result in appearance of a displaced vertex attributed to this sort of decay in addition to the production vertex (\ref{St-Nu-fin-3}).  The limit in (\ref{LIMIT0}) does not take into account such a possibility and sum up the event rates of sterile neutrino decay both within and outside a detector. 
However, one can imagine an experiment where the displaced vertices of the above mentioned type are looked for and are either observed or, more probably, excluded at certain confidence level.  For the latter case our limits in the region $m_{N}>$100 MeV would drastically change. In order to illustrate the influence of this additional criterium of event selection on our limits we impose on 
the processes (\ref{St-Nu-fin-3}) a condition that  sterile neutrino decays outside detector. This results in multiplication of the 
corresponding decay rate formulas  (\ref{Dec-TAU-PI-N}),  (\ref{DECAYRATETAU}) by the probability factor $1-P_{N}$. 
The modified limits take the form
 \begin{eqnarray}\label{LIMIT0-1}
|U_{\tau N}|^2& \leq& \mbox{Min} \left\{\frac{\Delta^{exp}(\tau^{-}\rightarrow \pi^{-} \nu)}{ \ \Gamma_\tau^{(\pi N) }},  \frac{\Delta^{exp}(\tau^{-}\rightarrow \nu \nu l^{-} )}{ \ \Gamma_\tau^{( N \nu l) }}\right\} \times \exp(L_{D} \Gamma^{0}_{N}).
\end{eqnarray}
Here we used an  inequality $\exp{(L_{D} \Gamma_{N})} \leq \exp{(L_{D} \Gamma^{0}_{N})}$, where 
$\Gamma_N^0 =  a_e(m_N)+  a_\mu(m_N) +  a_\tau(m_N)$ with $a_{e,\mu , \tau}$ 
defined in  (\ref{GT-comp-incl}), (\ref{coeff-a}).  In this case our exclusion curve  for $|U_{\tau N}|^{2}$ in Fig. 3 
in comparison to the case of (\ref{LIMIT0}) changes its part (e) to (d) leaving the part (c) intact.  
%
%in comparison to the case of (\ref{LIMIT0}) change its second part and becomes 
Now the exclusion curve (c)-(d) covers a mass region 
\mbox{$0 \leq m_{N}\leq m_{\tau}-m_{\pi} \approx 180$ MeV.}  Note again that this is just an illustration of  an impact of as yet non-existing experimental data allowing discrimination of the events with the displaced vertices associated with the sterile neutrino decay. 

%%%%%%%%%%%%%

  \begin{figure}[htbp]
\centering
\includegraphics[width=0.8\textwidth]{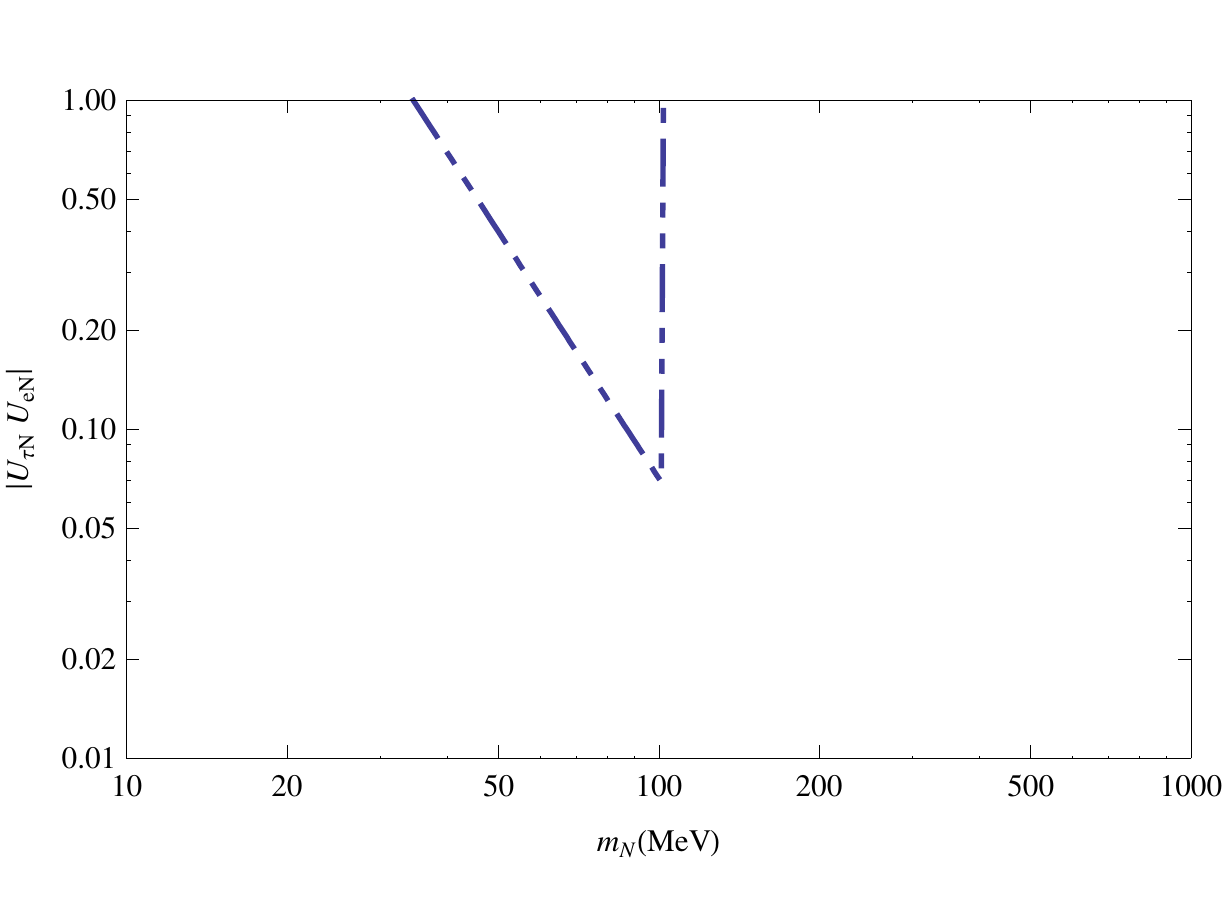}
\caption{ Exclusion curves for $|U_{\tau N} U_{e N}|$ from the present analysis.} 
\label{fig-4}
\end{figure}

  \begin{figure}[htbp]
\centering
\includegraphics[width=0.8\textwidth]{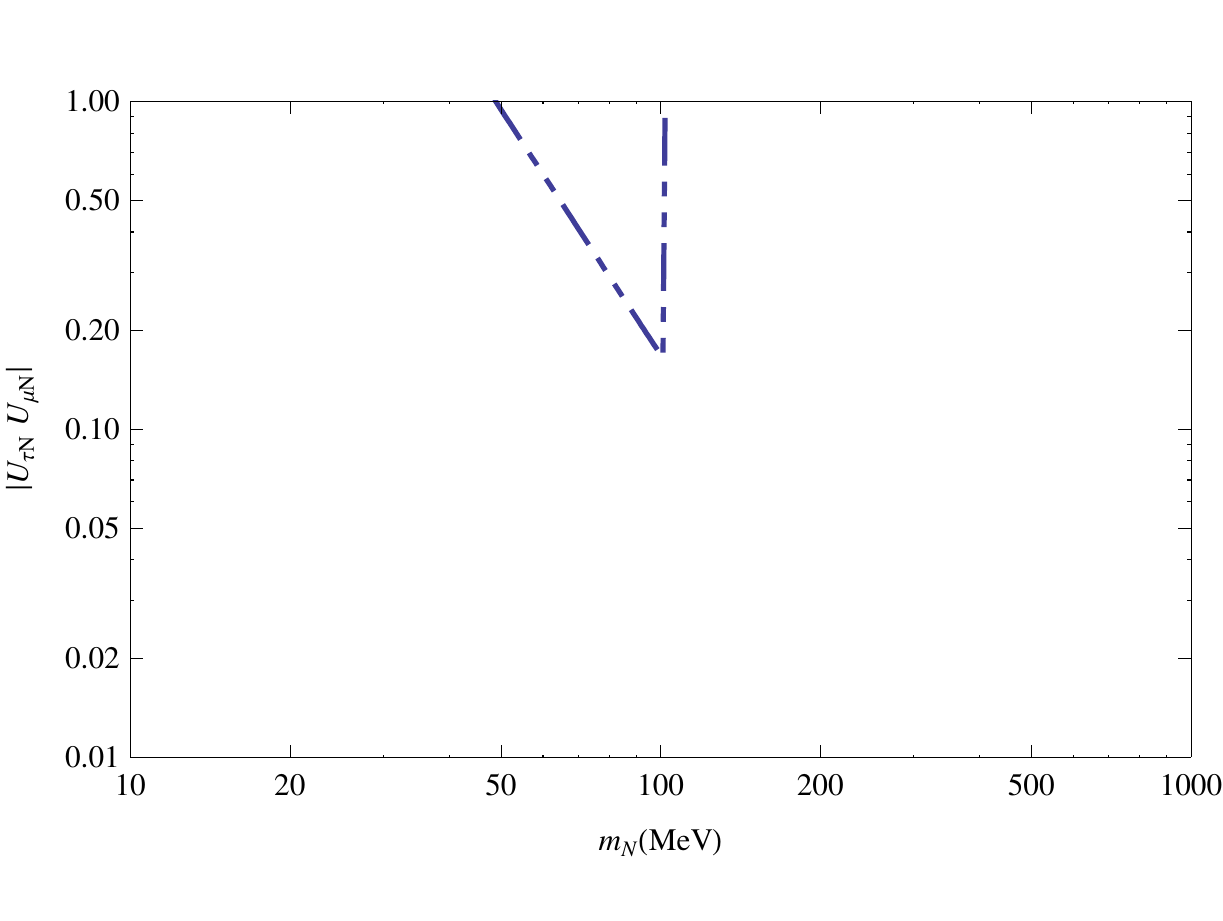}
\caption{Exclusion curves for $|U_{\tau N} U_{\mu N}|$ from the present analysis. } 
\label{fig-5}
\end{figure}

\section{Summary and Conclusions}

\label{sec-4}

We studied resonant contribution of sterile neutrino to leptonic and semileptonic decays of $\tau$ as well as to 
some semileptonic decays of $K$ and $D$ mesons.  Comparison of our predictions with the corresponding 
experimental data on these decays allowed us to extract new limits on the mixing matrix element $U_{\tau N}$
shown in Fig. \ref{fig-3} as curves (b), (c), (e).  In the two domains of the sterile neutrino mass 
0 $\leq m_{N}\leq 70$ MeV and 300 MeV $\leq m_{N} \leq$ 900 MeV our limits on $U_{\tau N}$ are 
more stringent than the limits existing in the literature.  For 0 $\leq m_{N} \leq$ 100 MeV our limit to a good approximation  is 
$|U_{\tau N}|^{2} \leq 2.9 \times 10^{-3}$.  We also obtained new, although not stringent, limits 
on the products $|U_{\tau N} U_{e N}|$ and $|U_{\tau N} U_{\mu N}|$ shown in Figs. \ref{fig-4}, \ref{fig-5}.
To our best knowledge there do not exist in the literature the limits on these products of the mixing matrix elements for
$m_{N}\leq$ 100 MeV. 

Our limits derived from the experimental results (\ref{N-fin-state-1})-(\ref{N-fin-state-2})
are, to certain extent, conservative estimates. In fact, let us assume that in derivation of these experimental values were applied specific kinematical criteria for event selection, suppressing possible contribution of $\tau\rightarrow \pi N, \  l \nu N$-decays
with a massive neutral particle N, such as sterile neutrino, instead of the nearly massless neutrino. Then, taking properly  into account these criteria in derivation of limits on $|U_{\tau N}|$ would have to strengthen them in comparison with our limits in Fig. \ref{fig-3}. This sort of analysis is beyond the scope of the present paper and  requires many additional and unknown for us details on the derivation of  
(\ref{N-fin-state-1})-(\ref{N-fin-state-2}) carried out by the corresponding experimental groups.  

We consider as an important point of our analysis  its model independent character in the sense that we do not refer to 
any sort of ad hoc assumptions about other two mixing matrix elements $U_{e N}$ and $U_{\mu N}$.  Such assumptions are typical 
for the existing literature on this subject. In particular, the limits of CHARM \cite{CHARM}   and 
NOMAD \cite{NOMAD} collaborations shown in Fig. \ref{fig-3} were obtained under the assumption 
$|U_{\tau N}|\gg |U_{\mu N}|, |U_{e N}|$.  At first site this assumption looks reasonable since the existing limits on 
$|U_{\mu N}|$ and $|U_{e N}|$ are very stringent (see, for instance, Ref. \cite{Atre:2009rg}). 
However, they were also obtained  under the assumptions of this type. To our mind  these observations should be taken into account  in  assessment of the  limits on the sterile neutrino mixing matrix elements $U_{\alpha N}$. 
%with active neutrino flavors $\nu_{\alpha}$. 
In some cases these limits may be rather stringent mainly because of this sort of assumptions.

\begin{acknowledgments}
We thank S. Kuleshov, G. Cvetic and C. Dib  for useful discussions.
This work was supported by FONDECYT projects 1100582 and 110287, and 
Centro-Cient\'\i fico-Tecnol\'{o}gico de Valpara\'\i so PBCT ACT-028. 
\end{acknowledgments}

\appendix*\section{Partial decay rates}

Here we specify the partial decay rates involved in Eqs. (\ref{DECRATEM})-(\ref{teta2}). For more details and discussion we refer reader to Refs.  
\cite{tau-decPaper,M-decPaper,Atre:2009rg} 

{\it The decay rates of mesons and $\tau$ to the final states with sterile neutrino} $N$:
 \begin{eqnarray}\label{DECAYRATETAU2}
%
%:{SK_4}  Esthetic change
\Gamma(M^{+}\rightarrow l_{i}^{+} N)&=&|U_{i N}|^{2} \frac{G_{F}^{2}}{8\pi} f_{M}^{2} |V_{M}|^{2} m_{M}^{3}
\lambda^{\frac{1}{2}}(x_{i}^{2},x_{N}^{2},1)(x_{i}^{2}+x_{N}^{2}-(x_{i}^{2}-x_{N}^{2})^2) \equiv \
\  |U_{i N}|^{2}\Gamma_M^{(l_i N)},\\
%{SK_4} END
\label{Dec-TAU-PI-N}
\Gamma( \tau^- \rightarrow  \pi^{-} N) &=& |U_{\tau N}|^2
\frac{G_F^2}{16 \pi}m_\tau^3 f_{\pi}^2 |V_{ud}|^{2} F_P(z_{N},z_{P})\equiv |U_{\tau N}|^2 \Gamma_\tau^{(\pi N)}, \\ 
\label{DECAYRATETAU}  
\Gamma(\tau^- \rightarrow l^{-}\nu_{l} N )&=& |U_{\tau N}|^2
\frac{G_F^2}{192\pi^3} m_\tau^5 I_{1}(z_{N},z_{\nu_{l}}, z_{l}) 
\equiv |U_{\tau N}|^2 \Gamma_\tau^{(l \nu N)},  \\ 
\label{DEC-TAU-3}
\Gamma(\tau^- \rightarrow l^{-}\nu_{\tau} N )&=& |U_{l N}|^2
\frac{G_F^2}{192\pi^3} m_\tau^5 I_{1}(z_{N},z_{\nu_{\tau}}, z_{l}) 
\equiv |U_{l N}|^2 \Gamma_\tau^{(l \nu N)}. 
\end{eqnarray}
Here we denoted $z_{i} = m_{i}/m_{\tau}$, $x_{i} = m_{i}/m_{M}$ with $m_{i} = m_{N}, m_{P}, m_l$. The kinematical functions $F_P(x,y)$, $I_1(x,y,z)$ are defined in \rf{kin-fun-1}.

{\it The partial decay rates heavy sterile neutrino}, $N$ including leptonic and semileptonic decay modes.   In the latter case the final hadronic states for low neutrino masses $m_{N}< m_{\rho}$ is represented by the lightest mesons while for larger 
$m_{N}>m_{\rho}$ by $q\bar{q}$-pairs as suggested by Bloom-Gilman duality \cite{BloomGilman}. 
This inclusive approach \cite{tau-decPaper} allows one to reduce uncertainties in the leptonic decay constants $f_{M}$ of  mesons starting from $\rho$-meson, some of which are only known in phenomenological models (for more details see \cite{tau-decPaper}).  The list of the sterile neutrino decay rates is as follows:
\begin{eqnarray}\label{lln-CC}
%
%%%%%%%%%%
%
\Gamma(N\rightarrow l_1^{-}l_2^{+}\nu_{l_{2}} )&=& |U_{l_1 N}|^2
\frac{G_F^2}{192\pi^3} m_N^5 I_{1}(y_{l_1},y_{\nu_{l_{2}}}, y_{l_2})(1-\delta_{l_{1}l_{2}}) 
\equiv |U_{l_1 N}|^2 \Gamma^{(l_1l_2\nu)}, \\
\label{lln}
\Gamma(N\rightarrow \nu_{l_{1}}l_2^{-}l_2^{+} )&=& |U_{l_1 N}|^2
\frac{G_F^2}{96\pi^3} m_N^5
\left[\left(g^{l}_{L} g^{l}_{R}+ \delta_{l_{1}l_{2}}g^{l}_{R}\right) I_{2}(y_{\nu_{l_{1}}}, y_{l_{2}}, y_{l_{2}}) + \right. \\ \nn
&&\left.   + \left((g^{l}_{L})^{2} +(g^{l}_{R})^{2 }+ \delta_{l_{1}l_{2}} (1 +2 g^{l}_{L})\right) I_{1}(y_{\nu_{l_{1}}}, y_{l_{2}}, y_{l_{2}}) \right] \equiv
\\ \nn
&\equiv& |U_{l_1 N}|^2\Gamma^{(l_2l_2\nu)},\\
\label{3n}
\sum_{l_{2}=e,\mu,\tau}\Gamma(N\rightarrow \nu_{l_{1}} \nu_{l_{2}} \bar{\nu}_{l_{2}})&=& |U_{l_1 N}|^2 
\frac{G_F^2}{96\pi^3} m_N^5  \equiv |U_{l_1 N}|^2
\Gamma^{(3\nu)},\\
\label{lP}
\Gamma(N\rightarrow l^{-}_{1} P^{+}) &=& |U_{l_{1}N}|^2
\frac{G_F^2}{16 \pi}m_N^3 f_{P}^2 |V_{P}|^{2} F_P(y_{l_{1}},y_{P})\equiv |U_{l_{1}N}|^2
\Gamma^{(lP)},\\
\label{nuP}
\Gamma(N\rightarrow \nu_{l_{1}} P^0) &=& |U_{l_{1}N}|^2 \frac{G_F^2}{64 \pi}m_N^3  f_{P}^2 
(1 - y^2_{P})^2 \equiv |U_{l_{1}N}|^2 \Gamma^{(\nu P)},\\ 
\label{lud}
%%%%%%%%%%
\Gamma(N\to l_{1}^{-} u\bar{d})&=&|U_{l_{1}N}|^2\ |V^{CKM}_{u d}|^{2} \frac{G_F^2}{64\pi^3}m_N^5 
I_{1}(y_{l_{1}}, y_{u}, y_{d})\equiv |U_{l_{1}N}|^2\Gamma^{(lud)},\\
 \label{nuqq} 
\Gamma(N\to \nu_{l_{1}}\, q\bar{q}) &=& |U_{l_{1}N}|^2\frac{G_F^2}{32\pi^3}m_N^5 \left[ g^{l}_{L} g^{l}_{R} I_{2}(y_{\nu_{l_{1}}}, y_{q}, y_{q}) + \right. \\ \nn
&&\left. + \left((g^{l}_{L})^{2} +(g^{l}_{R})^{2 })\right) I_{1}(y_{\nu_{l_{1}}}, y_{q}, y_{q}) \right] \equiv |U_{l_1 N}|^2 
\Gamma^{(\nu qq)}.
\end{eqnarray} 
Here $P=\pi, K$.  The decay constants are $f_{\pi}=130$MeV, $f_{K} = 159$MeV. 
We denoted $y_{i} = m_{i}/m_{N}$ with $m_{i} = m_{l}, m_{P}, m_{q}$. The CKM factors in Eq. (\ref{lP}) is 
$V_{\pi} = V^{CKM}_{ud}$, $V_{K} = V^{CKM}_{us}$. For the quark masses we use the values 
\mbox{$m_{u}\approx m_{d} = 3.5$ MeV},  \mbox{$m_{s} = 105$ MeV}, \mbox{$m_{c} = 1.27$ GeV}, \mbox{$m_{b} = 4.2$ GeV.}  In Eqs. (\ref{lud}), (\ref{nuqq})  we denoted  $u=u,c,t$; $d=d,s,b$ and $q=u,d,c,s,b,t$. The SM neutral current couplings of leptons and quarks are
\begin{eqnarray}\label{NC-coupl}
g^{l}_{L} &=&-1/2 + \sin^2\theta_W, \  \  g^{u}_{L}= 1/2 - (2/3) \sin^2\theta_W,  \ \ 
 g^{d}_{L}= -1/2 + (1/3) \sin^2\theta_W, \\
 \nn
g^{l}_{R} &=& \sin^2\theta_W,\ \ \ \ \ \ \ \ \ \ \  \ \,  g^{u}_{R}= -(2/3)\sin^2\theta_W,\  \ \ \ \  \ \ \,  g^{d}_{R}= (1/3)\sin^2\theta_W,
\end{eqnarray}

The kinematical functions in Eqs. (\ref{DECAYRATETAU2})-(\ref{nuqq}) are
\ba{kin-fun-1}
 &&I_{1}(x,y,z)= 12 \int\limits_{(x+y)^{2}}^{(1-z)^{2}} \frac{ds}{s}
(s-x^2-y^{2})(1+z^2-s) \lambda^{1/2}(s, x^{2}, y^2) \lambda^{1/2}(1, s, z^2),
\\ 
&& I_{2}(x,y,z)= 24 y z \int\limits_{(y+z)^{2}}^{(1-x)^{2}} \frac{d s}{s} (1+x^{2}-s)
\lambda^{1/2}(s,y^{2},z^{2})\lambda^{1/2}(1,s,x^{2}), \\ 
 \label{FP-1}
&&F_P(x,y)= \lambda^{1/2}(1,x^2,y^2) [(1+x^2)(1+x^2-y^2) - 4 x^2].
%
%&&F_V(x,y)= \lambda^{1/2}(1,x^2,y^2) [(1-x^2)^2+(1+x^2)y^2 - 2 y^4],\\ \nn
%
\ea

\begin{figure}[htbp]
\centering
\includegraphics[width=0.8\textwidth]{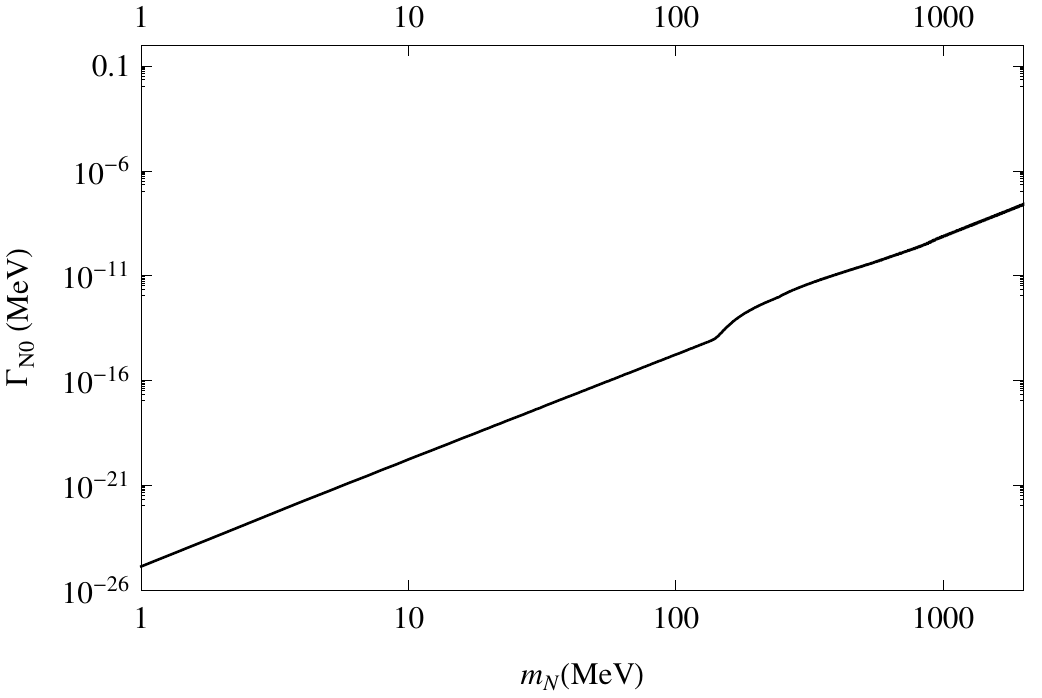}
\caption{Sterile neutrino decay rate $\Gamma_{N}$ for the particular case of 
$U_{e N}=U_{\mu N}=U_{\tau N}=1$ denoted by $\Gamma_{N0}$.}
\label{fig-6}
\end{figure}

The total decay rate $\Gamma_{N}$  of the heavy neutrino $N$ is equal to the sum of the partial decay rates in Eqs. (\ref{lln-CC})-(\ref{nuqq}),
which we write in the form: 
%:{SK_5}  Think on Majorana factors
%
\ba{total-4}
\Gamma_{N}& =&  \sum_{l_{1}, l_{2}, {\cal H}} (1+\delta_{N}) \left[  \Gamma(N\rightarrow l_{1}^{-}{\cal H}^{+}) +
  \Gamma(N\rightarrow l_{1}^{-} l_{2}^{+}\nu_{l_{2}}) +\right. \\ \nn
&+&\left.\Gamma(N\rightarrow \nu_{l_{1}}{\cal H}^{0}) + \Gamma(N\rightarrow l_{2}^{-} l_{2}^{+}\nu_{l_{1}}) + 
\Gamma(N\rightarrow \nu_{l_{1}}\nu_{l_{2}} \bar{\nu}_{l_{2}}) \right],
\ea
where we denoted the hadronic states ${\cal H}^{+} = P^{+}, \bar{d} u,  \bar{s} u,  \bar{d} c,  \bar{s} c$ and  
${\cal H}^{0} = P^{0}, \bar{q} q$. We introduced the factor $\delta_{N} = 1$ for Majorana and $\delta_{N}=0$ for Dirac neutrino $N$. Its appearance  is related with the fact that for Majorana neutrinos both charge conjugate final states are allowed: 
$N\rightarrow l_{1}^{-}l_{2}^{+} \nu_{l_{2}}, l_{1}^{+}l_{2}^{-} \bar{\nu}_{l_{2}}$; 
$N\rightarrow l_{2}^{-}l_{2}^{+} \nu_{l_{1}}, l_{2}^{+}l_{2}^{-} \bar{\nu}_{l_{1}}$
and $N\rightarrow l^{\mp}{\cal H}^{\pm}$. 
%
%:{SK_0} This part should be coordinated with the previous section for the paper and the thesis.
%
For convenience we write Eq. \rf{total-4} in the form:
\begin{eqnarray}\label{GT-comp-incl}
\Gamma_{N} &=& a_{e}(m_{N})\cdot  |U_{eN}|^{2}  + a_{\mu}(m_{N})\cdot   |U_{\mu N}|^{2} +  
a_{\tau}(m_{N})\cdot   |U_{\tau N}|^{2}
\end{eqnarray}
where 
\begin{eqnarray}\label{coeff-a}  
a_{l}(m_{N}) = 
(1+\delta_{N})\left[\Gamma^{(l {\cal H})} + \Gamma^{(3\nu)} +\sum_{l_{2}} \left(\Gamma^{(l_{2}l_{2}\nu)} + \Gamma^{(l l_{2} \nu)}\right)\right],
\end{eqnarray}
with $l, l_{2} = e, \mu, \tau$. In the  inclusive approach the hadronic contribution is calculated as
\begin{eqnarray}\label{H-incl}
\Gamma^{(l{\cal H})} = \theta(\mu_{0}-m_{N})\sum_{P=\pi,K}\left(\Gamma^{(\nu P)}  +  \Gamma^{(lP)}\right) +
\theta(m_{N} - \mu_{0})\sum_{u,d,q} \left(\Gamma^{(lud)} + \Gamma^{(\nu qq)}\right)
\end{eqnarray}
The parameter $\mu_{0}$ denotes the mass threshold from which we start taking into account hadronic contributions via 
$q \bar{q}$ production. In Refs. \cite{tau-decPaper,M-decPaper} we have shown that the reasonable choice is  
$\mu_{0} = m_{\rho^+} = 775.8$ MeV, which we also use in the analysis of present paper. In Fig.\ref{fig-6} we plotted 
$\Gamma_{N0}\equiv \Gamma_{N}(U_{e N}=U_{\mu N}=U_{\tau N}=1)$ as a function of the sterile neutrino mass $m_{N}$.

\end{document}